\documentclass[%
 reprint,
 amsmath,amssymb,
 aps,
 pra,
]{revtex4-1}

\usepackage{graphicx}
\usepackage{dcolumn}
\usepackage{bm}
\usepackage{color}
\usepackage{times}

\begin{document}

\preprint{APS/123-QED}

\title{Experimental realization of phonon demultiplexing in three-dimensions}

 \author{Osama R. Bilal$^{1,2}$} 
 \author{Chern Hwee Yee$^{2}$, Jan Rys$^{2}$, Christian Schumacher$^{3}$, and Chiara Daraio$^{4}$}
 \affiliation{$^{1}$Department of Mechanical engineering, University of Connecticut, Storrs, USA}
 \affiliation{$^{2}$Department of Mechanical and Process engineering, ETH Z\"urich, 8092 Z\"urich, Switzerland}
 \affiliation{$^{3}$Computer Graphics Laboratory, ETH Z\"urich, 8092 Z\"urich, Switzerland}
 \affiliation{$^{4}$Division of Engineering and Applied Science, California Institute of Technology, Pasadena, California 91125, USA}

\date{\today}
\begin{abstract}
Phononic metamaterials enabled the realization of many acoustic components analogous to their electronic counterparts, such as transistors, logic gates and calculators. A key component among these is the demultiplexer, a device that receives multiple signals and sorts them based on their frequencies into separate channels. Previous experimental realizations of acoustic and elastic multiplexers have employed plates with pillars or holes to demultiplex frequencies. However, existing realizations are confined to two-dimensions, which can limit potential acoustic or elastic circuit design. Here we show the first experimental realization of a three-dimensional, four	channel phononic demultiplexer. Our design methodology is based on bundles of pass-bands within a large band gap that can easily be tuned for multi-channel frequency demultiplexing. The proposed design can be utilized in acoustic and	elastic information processing, nondestructive evaluation and communication applications among others.
\end{abstract}

\maketitle

Phononic crystals and acoustic metamaterials consist of periodic arrangements of basic building blocks that repeat in space. These materials have the ability to control waves in an unprecedented manner, at different length and frequency scales. One of the unique properties of these materials is their ability to support forbidden frequency bands (i.e., band gaps) in their dispersion spectrum. Within a band gap, waves can not propagate and are reflected towards the source. The main mechanisms for creating these forbidden bands within the frequency spectrum are either destructive interference  (e.g., in Bragg scattering band gaps) or local resonance. When a band gap is open due to Bragg scattering, the wavelength of the affected waves is usually at the same order of the spacing between the unit cells (i.e., lattice spatial periodicity) \cite{kushwaha1993acoustic,sigalas1993band}. In contrast, when a band gap is open due to resonances, the wavelength of the attenuated waves can be independent of the lattice spacing \cite{liu2000locally}. Such resonances induce properties that might not exist in conventional materials, like negative effective mass or stiffness \cite{christensen2015vibrant,cummer2016controlling,ma2016acoustic}. Phononic crystals and metamaterials with such exotic properties have been utilized for many applications such as vibration and sound insulation \cite{yang2010acoustic,bilal2018architected}, seismic wave protection \cite{kim2012seismic,brule2014experiments}, wave guiding \cite{Torres_1999,rupp2007design} and frequency filtering \cite{pennec2004tunable,rupp2010switchable} among others \cite{maldovan2013sound}.

\begin{figure}[b]
	\begin{center}
		\includegraphics[scale =1]{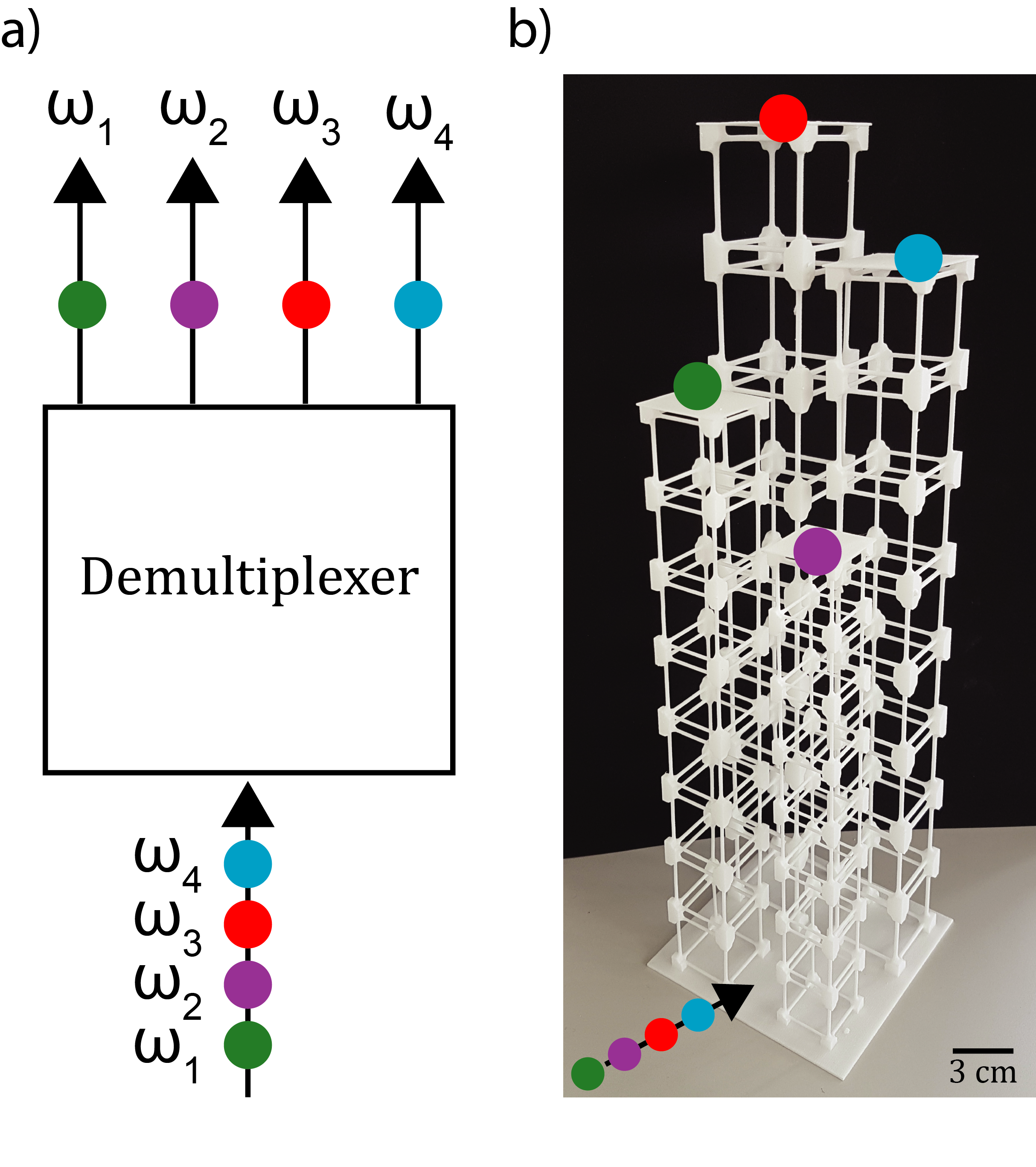}
	\end{center}
	\caption{ \textbf{3D Phononic demultiplexer}. a) A conceptual schematic of a 4-channel demultiplexer with four different input frequencies excited at one end and sorted in four predefined channels at the other end. b) A 3D printed realization of the 4-channel phonon demultiplexer, where four different frequencies are excited at the base and passively sorted into their corresponding channel.}
	\label{fig:Schematic}
\end{figure}

Phononic crystals and metamaterials have been proposed as platforms to enable mechanical information processing. Potential applications range from thermal computing \cite{li2006negative, wang2007thermal,joulain2016quantum} (at small scales) to ultrasound and acoustic based computing (at larger scales) \cite{bringuier2011phase,zhang2015acoustic}. Current realizations of fundamental phonon computing elements such as acoustic switches \cite{babaee2015harnessing,bilal2017reprogrammable}, rectifiers \cite{liang2009acoustic,liang2010acoustic}, diodes \cite{popa2014non,li2011tunable,devaux2015asymmetric,boechler2011bifurcation}, transistors \cite{hatanaka2013phonon,bilal2017bistable}, memory \cite{hatanaka2014mechanical} and lasers \cite{vahala2009phonon,jing2014pt} have been inspired by their electronic counterparts. Among such devices, demultiplexers are a combinational logic devices that take a single input at one end and route it to one of several output channels (Fig. \ref{fig:Schematic}a). There exist multiple theoretical proposals for realizing phonon demultiplexers for both acoustic  \cite{rostami2016designing, pennec2004tunable,watkins2020demultiplexing} and elastic waves \cite{zou2017decoupling, moradi2019three, motaei2020eight,nazari2020heterostructure,ben2020shaped, gharibi2020phononic, pennec2005channel, hussein2005hierarchical, vasseur2011band} with few experimental demonstrations \cite{mohammadi2011chip,faiz2020experimental,watkins2020demultiplexing}. The main working principle for most of these designs is based on embedded cylinders \cite{pennec2004tunable,vasseur2011band,rostami2016acoustic, zou2017decoupling}, holes \cite{mohammadi2011chip} or pillars \cite{faiz2020experimental} in a host plate. These realizations constrain demultiplexing phonons to two-dimensions. A device that can demultiplex phonons in three dimensions for all wave polarization remains elusive. In this paper, we numerically design and experimentally characterize a 3D phonon demultiplexer (Fig. \ref{fig:Schematic}b). 



\begin{figure}[b]
	\begin{center}
		\includegraphics[scale = 1] {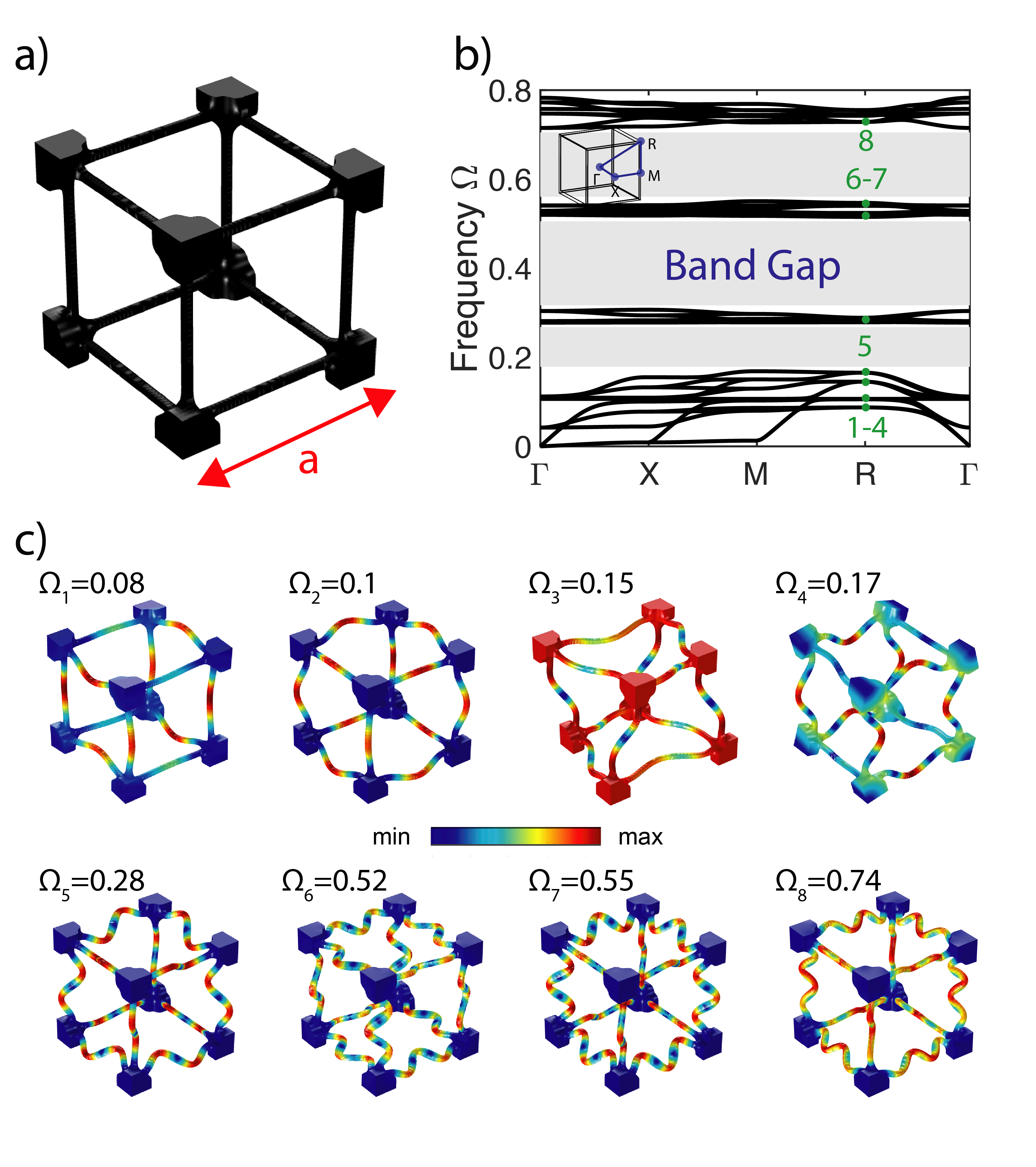}
	\end{center}
	\caption{\textbf{Basic building block.} a) The unit cell composed of eight masses at its corners connected with 12 beams forming a cube. b) Dispersion curves of the metamaterial unit cell along the path $\Gamma-X-M-R-\Gamma$ assuming infinite repetition of the unit cell in all directions. Full band gaps, where waves are prohibited from propagation, are highlighted in gray. The inset shows the irreducible Brillouin zone for a symmetric cubic unit cell. (c) Selected elastic mode shapes of the unit cell at the $R$ point at the four different transmission bands.}
	\label{fig:Unit_cell}
\end{figure}

To realize our demultiplexer, we design a cubic unit cell that has two important characteristics: (1) a wide band gap to filter out undesirable frequencies and (2) a bundle of pass bands inside the large band gap to allow only the targeted frequencies to pass. The unit cell is composed of eight masses, one at each of the unit cell's corners, connected by 12 beams~(Fig.~\ref{fig:Unit_cell}a).

To verify our design hypothesis, we first consider an infinite unit cell model, where one single unit cell is assumed to repeat in space in all directions. We analyze the unit cell using Bloch periodic boundary conditions \cite{bloch1929quantenmechanik}. The dispersion curves of the unit cell, correlating wavenumber to frequency, are calculated using the wave equations for heterogeneous media \cite{graff2012wave} within an infinite medium. We use the finite element method to solve the elastic wave equations (COMSOL 5.4). The solution is the wavefunction $u (x, \kappa; t) = \tilde{u} (x) \exp^{ (i (\kappa^\intercal x-\omega t))}$, where $\tilde u$ is the Bloch displacement vector, $x$ is the position vector, $\kappa$ is the wavenumber, $\omega$ is frequency and $t$ is time. The resulting dispersion curves are normalized by the unit cell size and the speed of sound in the medium, $\Omega = f a/c$. The dispersion plot shows a significant band gap (shaded gray region in Fig. \ref{fig:Unit_cell}b) with relative width of $\approx 124\%$. The relative width of the band gap is calculated as the absolute width of the band gap divided by its central frequency. Within the band gap region, there exist two pass bands with central frequency $\Omega = 0.29~\&~0.53$, which we engineer to work as the transmission bands for our demultiplexer.

To better explain the behavior of the proposed design, we visualize the unit-cell's vibrational mode shapes by superimposing the displacement profiles as a heat map over its geometry for eight different frequencies (Fig. (\ref{fig:Unit_cell}c)). The first two mode shapes ($\Omega_{1,2} = 0.08~\&~0.1$) resemble the first vibrational modes of the beams connecting the corner masses. The third and fourth modes ($\Omega_{3,4} = 0.15~\&~0.17$) show the mixed resonance modes of the corner masses along with the first vibrational modes of the connecting beams. All four modes exist below the the band gap. The fifth mode shape, which exists in the first pass band ($\Omega_{5} = 0.28$), shows the second vibrational mode of the connecting beams. The third vibrational mode of the unit cell beams manifests itself within the second pass band ($\Omega_{6,7} = 0.52~\&~0.55$). At the edge of the band gap ($\Omega_{8} = 0.74$), the connecting beams vibrate at their fourth mode shape. The masses work as pivots or fixation points for the connecting slender beams, which behave as fixed-fixed beams with well defined vibrational mode shapes. Based on this simple design concept, the unit cell can be considered as a simple mass spring model with a large band gap with embedded bands of transmission. By changing the diameter of the beams or the ratio between the mass of the beam to the mass of the corner block, we can change the position of the band gap and its width. Alternatively, by scaling the unit cell length $a$, we can shift the position of the band gap along with its embedded pass bands within the frequency spectrum. An advantage of scaling the entire unit cell while keeping all aspect ratios the same, is preserving all the dispersion characteristics of the unit cell (e.g., group velocities) without alteration. This preservation can help in maintaining consistency in the performance of the different channels, such as the mode shapes and wave polarization.

\begin{figure*}
	\begin{center}
		\includegraphics{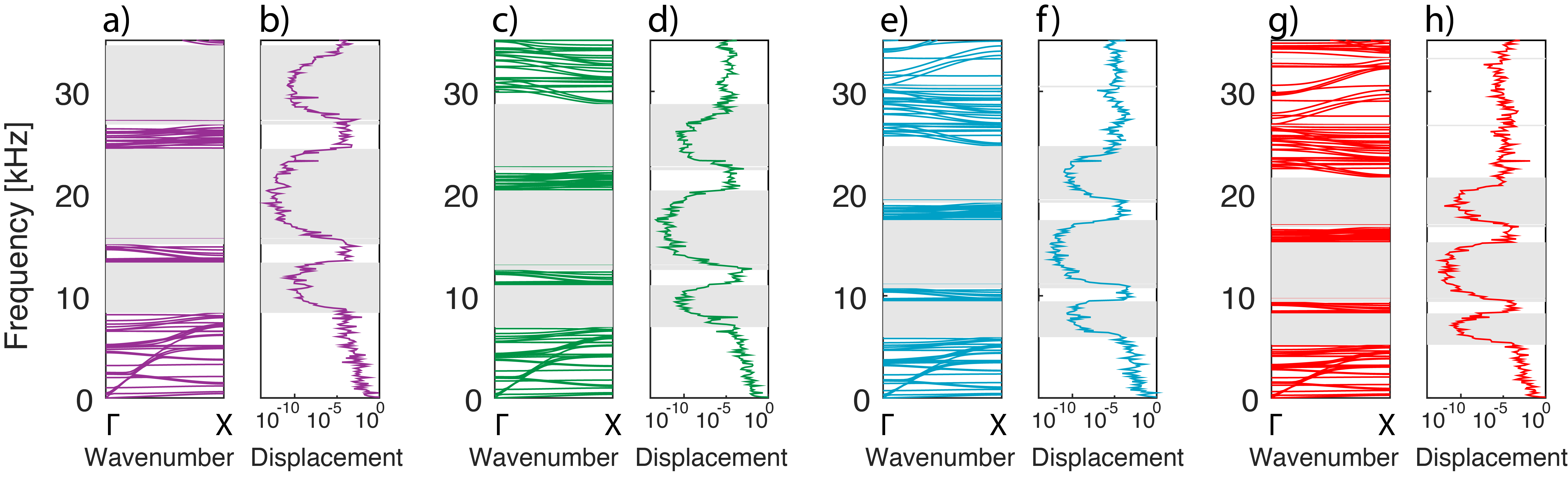}
	\end{center}
	\caption{\textbf{Infinite medium vs finite structure.} The calculated dispersion curves and their corresponding frequency response functions (FRF) for unit cells with side length equal to (a-b) 30 mm (c-d) 36 mm (e-f) 42 mm (g-h) 48 mm. The dispersion curves are calculated along the $\Gamma-X$ direction with periodicity only along the x direction. The FRFs are calculated for a structure composed of $1 \times 1  \times 8$ unit cells. The band gaps are highlighted in gray in both dispersion curves and FRFs.}
	\label{fig:Disp_FRF}
\end{figure*}


The demultiplexer is constructed by designing four channels out of the $same$ unit cell, but with four different lattice constants. All four channels are connected at one end (i.e, the base of the demultiplexer). The excitation takes place at the base with multiple frequencies that get sorted into the four channels. To correctly confirm the behavior of the scaled unit cells, we calculate the dispersion curves for four different lattice constants $a = $ 30, 36, 42 and 48 mm (Fig. \ref{fig:Disp_FRF}a,c,e and g). The dispersion curves for these four unit cells are calculated based on 1D Bloch periodicity (only along the $\kappa_x$ direction). Basing the analysis on 1D periodicity, as opposed to 3D periodicity, gives rise to extra modes along the free boundaries with no periodicity. These boundary modes manifest themselves as added lines in the dispersion curves. A clear difference between the 1D and 3D periodic dispersion curves can be seen, for example, by comparing the number of branches below the first band gap in Fig. \ref{fig:Unit_cell}b and \ref{fig:Disp_FRF}a. The 1D periodic dispersion curves serve as the limiting case for the performance of the proposed design. As more periodicity is incorporated in the model (i.e., along $\kappa_y ~\& ~\kappa_z$ direction), less vibration modes become available at the free boundaries and the dispersion curves become identical to figure \ref{fig:Unit_cell}b. Therefore, having a demultiplexer channel with more unit cells in all directions will result in less dispersion lines and wider band gaps.

\begin{figure}[b]
	\begin{center}
		\includegraphics[width =\columnwidth]{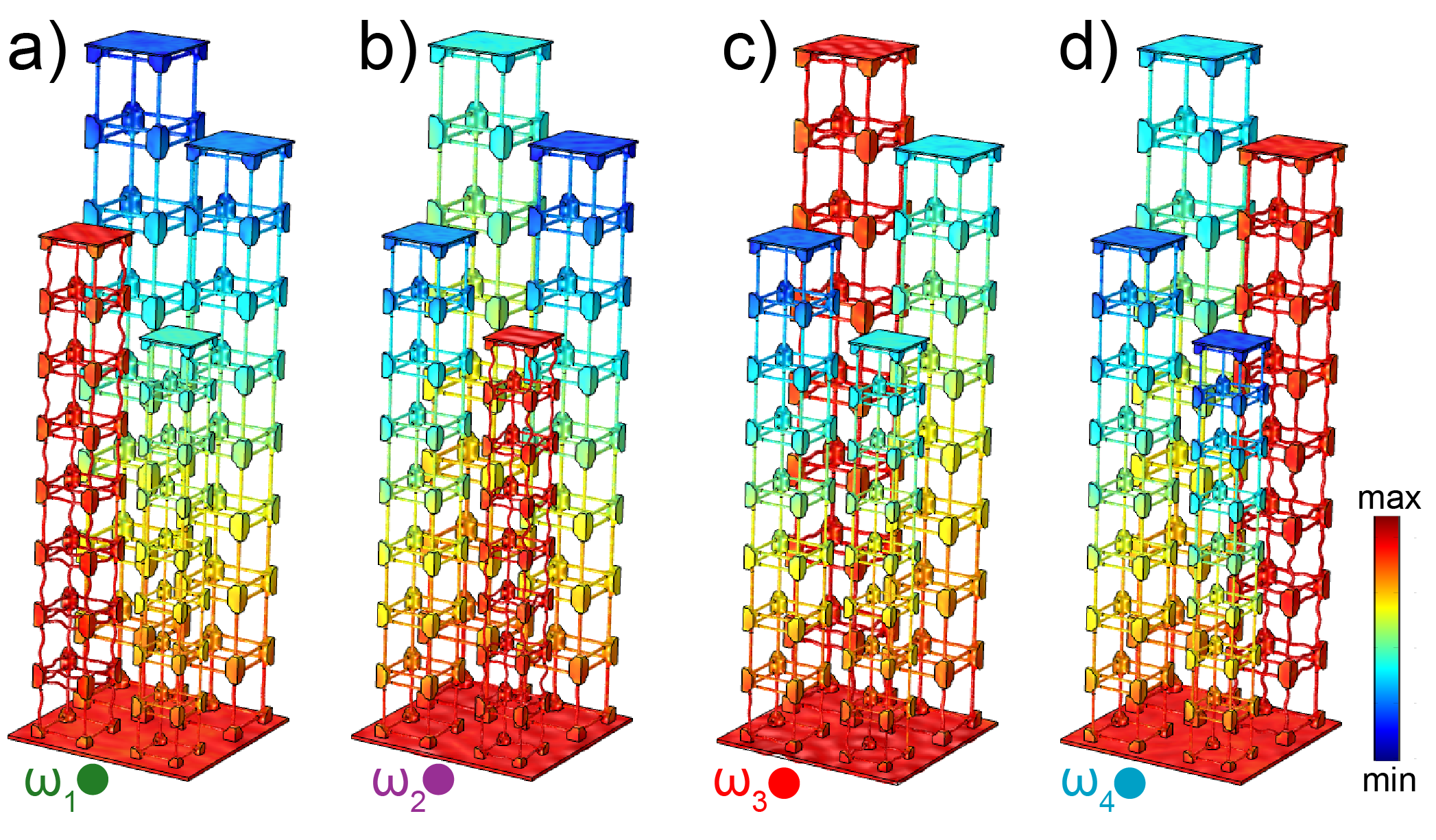}
	\end{center}
	\caption{\textbf{Demultiplexer numerical simulations.} Numerically
		calculated mode shapes of the demultiplexer operating at a) $\omega_1$ = 11.7 kHz, b) $\omega_2$ = 13.9 kHz, c) $\omega_3$ = 15.9 kHz and d) $\omega_4$ = 18.1 kHz. The heat map represents the intensity of motion (i.e., the logarithm of displacement) through the entire structure when excited at the base.}
	\label{fig:Mode_shapes}
\end{figure}
\begin{figure}[b]
  \begin{center}
    \includegraphics{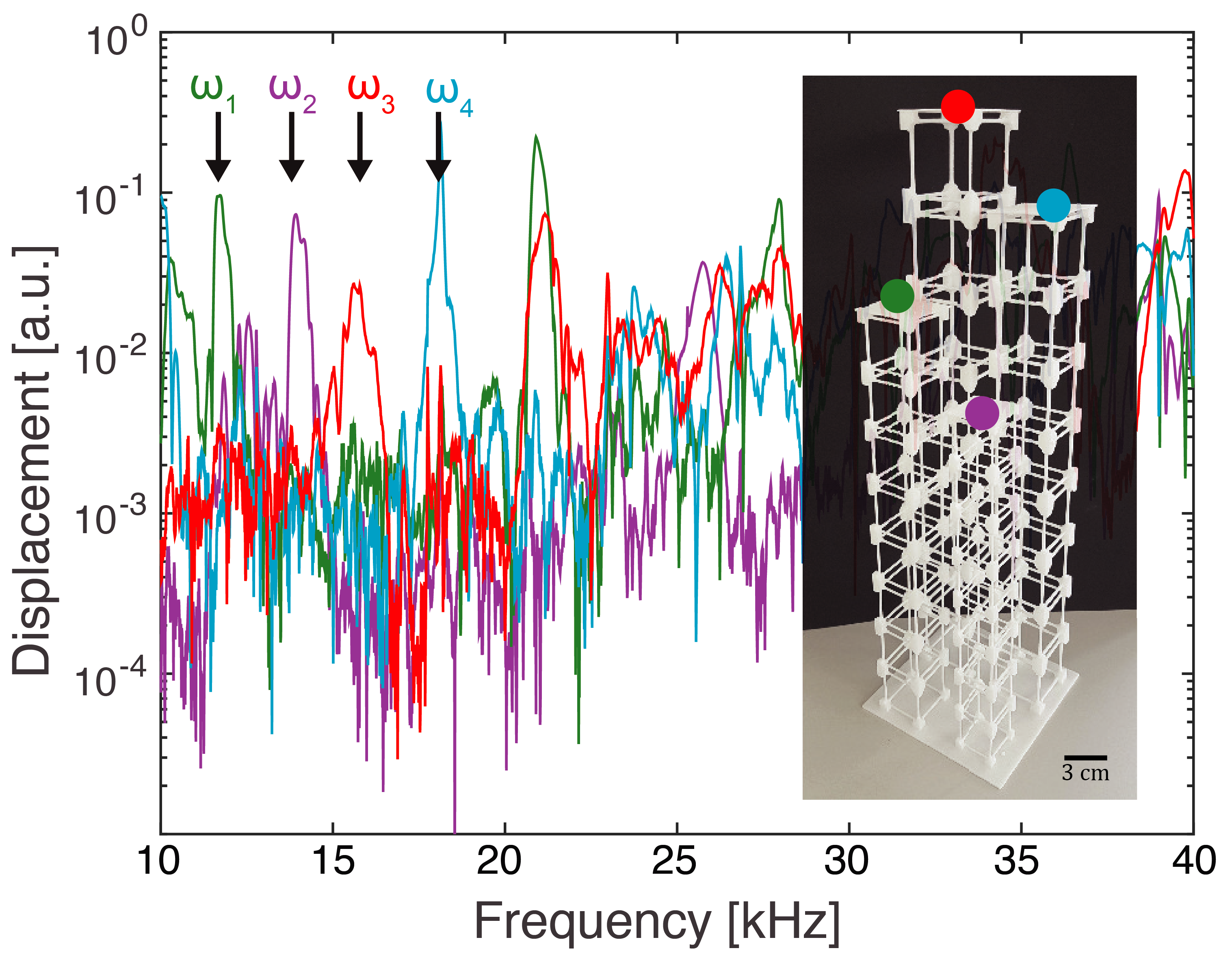}
  \end{center}
  \caption{\textbf{Demultiplexer experimental characterization.} Experimental measurements of the displacement at the end of each channel as a function of frequency. The targeted frequencies are marked with black arrows with $\omega_1$ = 11.7 kHz, $\omega_2$ = 13.9 kHz, $\omega_3$ = 15.9 kHz and $\omega_4$ = 18.1 kHz.}
  \label{fig:Experiments}
\end{figure}
To validate the infinite unit cell model against the finite structure with all the connected channels, we numerically simulate the exact geometry of the demultiplexer as seen in figure \ref{fig:Schematic}b using the finite element method. Each channel is composed of an array of 8 unit cells tessellated along the z direction. All channels are connected to a base plate with thickness $T_h$ = 3mm. The top end of each channel is terminated by a plate with the same thickness for signal extraction. The channels are arranged in a 2 $\times$ 2 grid separated by 30 mm. A harmonic excitation is applied at the bottom surface of the base plate. We sweep the excitation frequency between 1 and 32 kHz, and record the displacement at the end of each of the channels (Fig. \ref{fig:Disp_FRF}b,d,f and h). The predicted band gap frequencies using the infinite unit cell model are shaded in gray in all panels of figure \ref{fig:Disp_FRF}. The results of both the finite and infinite models are in a very good agreement. Within the band gaps, shown as the gray shaded regions in figure \ref{fig:Disp_FRF}, the amplitude of the transmitted wave is many orders of magnitude less than that of the pass band frequencies. The fact that all channels are connected at the base is not affecting the robustness of the individual channels' performance. In addition, we also plot the logarithm of the full displacement fields as a heat map at the four operational frequencies of the demultiplexer (Fig. \ref{fig:Mode_shapes}). The displacement field at each frequency shows a clear transmission of elastic phonons at targeted frequency, while the rest of the channels show negligible motion.

To experimentally validate our numerical simulations, we fabricate the demultiplexer through additive manufacturing (laser sintering) using polyamide (Young's modulus $E$ = 2.3 GPa and density $\rho$ = 1200 Kg/$m^{3}$). We characterize the vibration response of the meta-device by harmonically exciting the base plate with a piezoelectric disc (repeating the previously performed numerical simulations in figure \ref{fig:Disp_FRF}, \ref{fig:Mode_shapes}). We measure the transmitted vibrations at the end of each channel with a laser Doppler vibrometer LDV (Polytec OFV-505 with a OFV-5000 decoder, using a VD-06 decoder card). We excite the system by sweeping through frequencies between 10 and 40 kHz and record the amplitude of the transmitted vibrations at the end of each channel (Fig. \ref{fig:Experiments}). The measured vibrations between 10 and 20 kHz show a clear band gap, as predicted by the numerical simulations in all channels, and demonstrate clearly transmission of the signal in the designated pass bands in each channel. The experimental measurements show a clear evidence of phonon demultiplexing based on frequency in the realized meta-device. 

In this study, we design, analyze and realize the first 3D phononic demultiplexer. Our design relies on defining relatively narrow pass bands within a wide band gap. Designing the width of the pass band allows for a wider operational bandwidth and ensures transmission of each signal to its designated channel. The complete band gap for all directions and polarization (in-plane, out of plane and rotation) of the wave allows for phonon demultiplexing in three-dimensions. Multiplexing  elastic and acoustic waves in 3D can enable the design of higher dimensional all-phononic circuits taking an advantage of the recent progress in advanced manufacturing. Due to these advances in fabrication and the material invariant nature of the design principle, our meta-device can be easily scaled to operate at desired frequencies with no geometry alteration. For example, the same design fabricated at half the size would operate around twice the frequency (i.e., in the ultrasound regime). Our demultiplexer has possible applications in imaging, communications and phonon based information processing. 


The authors are very grateful for the fruitful discussions with Dr. Foehr and his help with the experimental setup.

\bibliography{Multiplexer}

\end{document}